\documentclass{emulateapj}
\usepackage{natbib}
\newcommand{\msun}{\,{\rm M_\odot}}
\newcommand {\be} {\begin{equation}}
\newcommand {\ee} {\end{equation}}
\def\spose#1{\hbox to 0pt{#1\hss}}
\def\lta{\mathrel{\spose{\lower 3pt\hbox{$\mathchar"218$}}
           \raise 2.0pt\hbox{$\mathchar"13C$}}}
\def\gta{\mathrel{\spose{\lower 3pt\hbox{$\mathchar"218$}}
           \raise 2.0pt\hbox{$\mathchar"13E$}}}
\begin{document}

\title{Black-Hole Spin and Galactic Morphology}
\author{Marta Volonteri\altaffilmark{1}, Marek Sikora\altaffilmark{2,3} \& Jean-Pierre
Lasota\altaffilmark{4,5}}

\email{martav@umich.edu, sikora@camk.edu.pl, lasota@iap.fr}

\altaffiltext{1}{Department of Astronomy, University of Michigan, 500
Church Street, Ann Arbor, MI, USA}
\altaffiltext{2}{Nicolaus Copernicus Astronomical Center, Bartycka 18, 00-716
Warszawa, Poland}
\altaffiltext{3}{Department of Physics and Astronomy, University of Kentucky, Lexington, KY, USA}
\altaffiltext{4}{Institut d'Astrophysique de Paris, UMR 7095 CNRS, Universit\'e
Pierre et Marie Curie, 98bis Bd Arago, 75014 Paris, France}
\altaffiltext{5}{Astronomical Observatory, Jagiellonian University, ul. Orla 171,
30-244 Krak\'ow, Poland}

\begin{abstract}
We investigate the conjecture by Sikora, Stawarz \& Lasota (2007)
that the observed AGN--radio-loudness bimodality can be explained by
the morphology~-~related bimodality of black-hole spin distribution
in the centers of galaxies: central black holes in giant elliptical
galaxies  may have (on average) much larger spins  than black holes in
spiral/disc galaxies. We study how accretion from a warped disc
influences the evolution of black hole spins and conclude that
within the cosmological framework, where the most massive BHs have
grown in mass via merger driven accretion, one indeed expects most
supermassive black holes in elliptical galaxies to have on average
higher spin than black holes in spiral galaxies, where random, small
accretion episodes (e.g. tidally disrupted stars, accretion of
molecular clouds) might have played a more important role.
\end{abstract}
\keywords{
cosmology: theory -- black holes -- galaxies: evolution -- quasars:
general}

\section{Introduction}
It has been known for many years that  the radio loudness of
AGN hosted by  disc galaxies is on average three orders of
magnitude lower than the radio loudness of AGN hosted by giant
ellipticals  (see Xu et al. 1999 and references therein). However,
as as shown by HST observations, such a galaxy morphology --
radio-loudness correspondence is  not ``one-to-one": both
radio-quiet and radio-loud very luminous quasars are hosted by
giant ellipticals (Floyd et al. 2004). On the other hand the
popular version of the so-called ``spin paradigm" asserts that
powerful relativistic jets  are produced in AGN with fast rotating
black holes (Blandford 1990), implying that BHs rotate slowly in
radio-quiet quasars, which represent the majority of quasars. 
However, such conjecture, at least in its basic interpretation,
is in conflict with the high average BH spin in quasars
deduced from the high average radiation efficiency of quasars
using the ``So{\l}tan argument" (So{\l}tan 1982; Wang et al. 2006
and references therein).

Parallel studies of radio-emission from X-ray binaries showed
that at high accretion rates production of jets is intermittent
(Gallo et al. 2003) and that this intermittency can be related
to transitions between two different  accretion modes (Livio et al.
2003). This inspired Ulvestad \& Ho (2001), Merloni et al. (2003),
Nipoti et al. (2005), and K\"ording et al. (2006)  to postulate the
existence of a similar intermittency of  jet production
in quasars and formulate an ``accretion paradigm" according to which
the radio-loudness is entirely related to the states of accretion
discs. However, Sikora et al. (2007) found that on the
radio-loudness -- Eddington-ratio plane AGN form two parallel
sequences that occurrence  of cannot be explained by the
``accretion-paradigm" (see also Terashima \& Wilson 2003; Chiaberge
et al. 2005; and Panessa et al. 2007). Sikora et al. (2007)
therefore proposed a revised version of the ``spin paradigm",
suggesting that giant elliptical galaxies host, on average, black
holes with spins larger than those hosted by spiral/disc galaxies.

This morphology-related radio dichotomy breaks down at high
accretion rates where the dominant fraction of luminous quasars
hosted by elliptical galaxies is radio quiet. This
radio-quietness occurs in quasars with high spin values. In such
systems with high accretion rates the intermittency is related to
the conditions of production of collimated jets, in agreement with
what is found in X-ray binaries, and with the ``So{\l}tan
argument". It should be emphasized that even if the
production of powerful relativistic jets is conditioned by the
presence of fast rotating BHs, it also depends on the accretion
rate and on the presence of disc MHD winds required to provide the
initial collimation of the central Poynting flux dominated outflow.

In this article we will examine under which condition the
cosmological evolution of BHs in galaxies may lead to low spins in
disc galaxies and high spins in more massive ellipticals.

%Recently Sikora, Stawarz \& Lasota (2006) found that Active Galactic
%Nuclei (AGN) form two, well separated, sequences on the
%radio-loudness -- Eddington-ratio plane, the upper one (radio-loud)
%being populated only by elliptical hosts, the disc galaxies being
%confined to the lower sequence. They interpret this fact in the
%framework of the (revised) spin paradigm concluding that nuclear
%black holes in ellipticals have (on average) much larger spins than
%black holes in spiral/disc galaxies. Such interpretation requires
%explaining why black holes in disc galaxies have spins lower than
%black holes hosted by ellipticals \footnote{Notice, however, that most of luminous quasars are
%radioquiet despite being hosted by elliptical galaxies. Within the
%revised spin paradigm this is explained by positing that during high
%accretion-rate events the formation of collimated jets is suppressed
%for most of time as it seems to be happening in ``microquasars".}.

%In addition to mass, spin is the only other parameter characterizing
%properties of astrophysical black-holes. The values of these two
%parameters reflect, although not in an unique way, their origin
%and evolution (Volonteri et al. 2005).

To put  our investigation in the relevant context we will first
recall why the value of black-hole's spin might be of fundamental
importance for relativistic jet launching. Assuming that
relativistic jets are powered by rotating black holes through the
Blandford-Znajek mechanism, Blandford (1990) suggested that the
efficiency of jet production is determined by the dimentionless
black hole spin, $\hat a \equiv J_h/J_{max}=c \, J_h/G \, M_{\rm
BH}^2$, where $J_h$ is the angular momentum of the black hole. If
true, this could explain  the very wide range of radio-loudness of
AGN that look very similar in many other aspects by attributing it
to a corresponding black-hole spin distribution. This so called
``spin paradigm" was explored by Wilson \& Colbert (1995), who
assumed that the black-hole spin evolution is determined mainly by
mergers. They claimed that mergers of black holes, following mergers
of galaxies, lead to a broad, `bottom-heavy' distribution of the
spin, consistent with a distribution of quasar radio-loudness.
However, this claim was challenged by Hughes \& Blandford (2003),
who showed that mergers cannot produce the required fraction of
black holes with high spins and concluded that accretion of matter
is essential in determining black-hole spins. In this case, however,
as noticed earlier by Moderski \& Sikora (1996) and Moderski, Sikora
\& Lasota (1998; hereafter MSL), one encounters the difficulty of
maintaining a sufficient number of black holes at the required low
spin, the spin-up by accretion discs being so efficient. MSL could
match the distribution of radio-loudness with the spin distribution
only by feeding holes with very small randomly oriented accretion
events,  i.e. by accretion events forming co-rotating and
counter-rotating discs with the same probability.

MSL also addressed the problem of the spin overflipping due to
the Bardeen-Petterson effect.  When an accretion disc does not
lie in the equatorial plane of the BH, that is, when the angular
momentum of the accretion disc is misaligned with respect to the
direction of $J_h$, the dragging of inertial frame causes a
precession that twists the disc plane due to the coupling of $J_h$
with the angular momentum of matter in the disc. The torque
tends to align the angular momentum of the matter in the disc
with that of the black hole, causing thus the inclination angle between the angular momentum vectors to decrease with decreasing distance from the BH, forcing the inner parts of the accretion disc to rotate in
the  equatorial plane of the BH (Bardeen \& Petterson 1975). 
Sustained accretion from a twisted disc would align the BH spin (and the innermost equatorial disc) with the angular momentum vector of the disc at large radii (Scheuer \& Feiler 1996). If the disc was initially counter-rotating with respect to the BH, a complete overflip would eventually occur, and then accretion of co-rotating material would act to spin up the BH (Bardeen 1970).

MSL concluded that the Bardeen-Petterson effect can be neglected
because the alignment time ($10^7$ years; Rees 1978) is  longer than
the duration of a single accretion event. Later, however, a series
of papers put into doubt the validity of Rees's estimate (Scheuer \&
Feiler 1996, Natarayan \& Pringle 1998).  This framework was
recently investigated by Volonteri et al. (2005) who argue that the
lifetime of quasars is long enough that angular momentum coupling
between black holes and accretion discs through the
Bardeen-Petterson effect effectively forces the innermost region of
accretion discs to align with black-hole spins (possibly through spin flips), and hence all AGN black-holes should have large spins.

Recently King et al. (2005) pointed out that under some conditions
the alignment torque can lead to disc-hole counter-alignment
reactivating the debate. The  counter-alignment process was
numerically simulated by Lodato \& Pringle (2006).

We here re-analyze the alignment problem in view
of all these latest results. We explore what are the likely outcomes
of accretion episodes that grow black holes along the cosmic
history, and determine under which conditions black holes in disc
galaxies end up having low spins.

\section{Assumptions}

The dynamics involving a spinning black hole accreting matter from a
thin disc whose angular momentum is not aligned with the spin axis
has been studied in a number of papers (e.g., Papaloizou \& Pringle
1983; Pringle 1992; Scheuer \& Feiler 1996; Natarajan \& Pringle
1998). A misaligned disc is subject to the Lense-Thirring
precession, which tends to align the inner parts of the disc with
the the angular momentum of the black hole. The outer regions of the
disc are initially inclined with respect to the hole's axis, with a
transition between alignment and misalignment occurring in between
at the so-called ``warp radius" (see below). The direction of the
angular momentum of the infalling material changes direction as it
passes through the warp. In our calculations we will assume that the
black-hole spin evolution is determined by accretion only. Volonteri
et al. (2005) have shown that BH mergers play a sub-dominant role in
the global spin evolution.

\subsection{Viscosities}

Despite many efforts  the problem of warped discs, especially in the
non-linear regime, has yet to be solved. Therefore the
characteristic scales of the problem are subject to several
uncertainties. The main quantity of interest is the ``warp radius"
$R_{\rm w}$ defined as the radius at which the timescale for radial
diffusion of the warp is comparable to the local
dragging-of-inertial frame (``Lense-Thirring" in the weak-field
approximation) time $(\hat a\, c\,R_s^2/R^3)^{-1}$ (Wilkins 1972). The
timescale for the warp radial diffusion can be written as \be
t_w\approx \frac{R_w^2}{\nu_2} \ee where $\nu_2$ is a viscosity
characterizing the warp propagation which can be different from the
accretion driving viscosity, $\nu_1$, which is responsible for the
transfer of the component of the angular momentum parallel to the
spin of the disc. The relation between $\nu_1$ and $\nu_2$ is the
main uncertainty of the problem, assuming of course that such
two-viscosity description is adequate at all. Describing $\nu_1$ by
the Shakura--Sunyaev parameter $\alpha$ one can show (Papaloizou \&
Pringle 1983) that the regime in which $H/R < \alpha \ll 1$ ($H$
being the disc thickness) one has $\nu_1/\nu_2\approx \alpha^2$. In
such a case the accretion time $t_{\rm acc}\approx R_w/\nu_1$ would
be much longer than the warp diffusion time $t_w$.

However, such a description can be questioned on several grounds.
First, is the  $\alpha \ll 1$ appropriate for high-rate accretion
onto AGN black-holes? There are no reliable estimates of this
parameter for AGN but  outbursts of LMXBs suggest that in hot
accretion discs $\alpha \gta 0.1$ (see e.g. Dubus et al. 2001). In
such a case $\nu_1$ is comparable to $\nu_2$ (Kumar \& Pringle
1985). Second, even if the two viscosities are different is
$t_w$ the relevant time for black-hole re-alignment? This is not clear
since this latter process is very dissipative and could be controlled
by accretion and not warp propagation.
%Not being able to decide what
%the correct answer is we consider below three cases: i)
%$\nu_1=\nu_2$, $\alpha=0.03$, ii) $\nu_1=\nu_2$, $\alpha=0.1$ and
%iii) $\nu_2 =\nu_1/\alpha^2$ for both values of $\alpha$ listed
%above.

\subsection{Relevant radii}

During the accretion process, the angular momentum of the disc at
the warp location sums up with that of the black hole, so the angle
between the angular momentum of the outer disc and the BH spin
changes. King et al. (2005) suggest that the condition of alignment
or counter-alignment can be expressed as a function of the angular
momenta of the hole and of the disc:  $J_h$ and $J_d$.  The
counter-alignment condition depends on the ratio $0.5\,J_d/J_h$, to
be compared with the cosine of the inclination angle, $\phi$.  If
$\cos \phi<-0.5\,J_d/J_h$, the counter-alignment condition is
satisfied. King et al. (2005), however,  leave the definition of
$J_d$ vague, indeed they suggest that $J_d$ is the angular momentum
of the disc inside a certain radius $R_J$ such that $J_d(R_J)=J_h$.
First, this is not a useful operational definition, because in this
case $\cos \phi<-0.5\,J_d/J_h=-0.5$, is a static condition, which
does not depend on the properties of the black hole or  of the
accretion disc. Second, matter contained within radius $R_J$ cannot
transfer all its angular momentum to $R_w$ but only a fraction
$\sqrt{R_w/R_J}\,J_d(R_J)$. Therefore a more natural radial scale in
the problem is  the warp radius $R_w$ and in the following we will
assume that $J_d= J_d(R_w)$ (note that we share this choice with
Lodato \& Pringle 2006).

\section{Method}

We explore the dependence of the alignment timescale in a
Shakura-Sunyaev disc on: viscosity $\nu_2/\nu_1$, black hole mass $M_{\rm BH},$
misalignment angle, Eddington ratio, accreted mass $m$. Some articles on this subject use the solution of Collin-Souffrin \& Dumont (1990), however in the view of the basic uncertainties of the problem we decided to use the less
refined solution of Shakura \& Sunyaev 1973.

Assuming a Shakura-Sunyaev disc  ({``}middle region"), the warp
radius (in units of the Schwarzschild radius $R_S$) can be expressed
as:

\begin{eqnarray}
\frac{R_w}{R_s  }=
3.6\times 10^3 \hat
a^{5/8}\left(\frac{M_{\rm BH}}{10^8 \msun}\right)^{1/8}\times  \nonumber \\
f_{\rm Edd}^{-1/4}\left(\frac{\nu_2}{\nu_1}\right)^{-5/8}\alpha^{-1/2}.
\label{eq:rw}
\end{eqnarray}
where  $f_{Edd} \equiv \dot M c^2 / L_{Edd}.$

We can then define the accretion timescale:

\begin{eqnarray}
t_{\rm acc}=\frac{R_w^2}{\nu_1}=3\times10^{6}{\rm yr}\, \alpha^{-3/2}
\left(\frac{\nu_2}{\nu_1}\right)^{-7/8}\,\hat a^{7/8}\times  \nonumber \\
f_{\rm Edd}^{-3/4}\, \left(\frac{M_{\rm BH}}{10^8\msun}\right)^{11/8}
\end{eqnarray}
(where $\nu_1= \alpha H^2 \Omega_K$ was used),
and the timescale for warp propagation:
\begin{equation}
t_w=\frac{\nu_1}{\nu_2}t_{\rm acc}.
\end{equation}
The ratio of angular momenta of the disc at $R_w$, defining
$M_d(R_w)=\dot M t_{\rm acc}(R_w)$, and of the black hole is:
\begin{eqnarray}
\frac{J_d}{J_h}=\frac{M_d}{\hat a M_{\rm
BH}}\left(\frac{R_w}{R_s}\right)^{1/2}=\nonumber \\
2\times10^{-9}f_{\rm Edd} \times 
\left({t_{\rm acc}\over {\rm
1\,y}}\right)\left(\frac{R_w}{R_s}\right)^{1/2} \hat a^{-1}.
\end{eqnarray}
%where we express the accretion rate in units of the Eddington rate:
%$M_d={\dot M} t_{acc}$, and $\dot M=f_{\rm Edd}\dot m_{\rm Edd}$.

%The warp radius depends only on BH mass and spin, being larger for
%larger masses and spins. The counter-alignment condition depends on the
%ratio $0.5\,J_d/J_h$, to be compared with the cosine of the inclination
%angle.

\section{Results}
\subsection{Single accretion episodes}

We first discuss  the behavior of the disc+BH system during the
alignment process.
The scheme we adopt is as follows:

\begin{enumerate}
    \item for a BH with initial mass $M_{\rm BH0}$ determine the initial conditions: warp radius, $R_w$,
timescale
for warp propagation
$t_w=R_w^2/\nu_2 $ ($\nu_2$ is chosen either coincident with $\nu_1$,
or $\nu_2=\nu_1/\alpha^2$),
accretion timescale for material at the warp radius,
$t_{\rm acc}=R_w^2/\nu_1$, angular momentum of the hole and of the disc at
$R_w$,  $J_h$ and $J_d$.

\item using the King et al. (2005) condition for misalignment
determine if the
BH and the inner disc are aligned or counter-aligned (counter-aligned if
$\cos
\phi < -0.5 J_d/J_h$).

\item over timesteps $\Delta t = t_{\rm acc}(R_w)$
compute the necessary quantities at the end of every step: increase
in black hole mass due to accretion, new BH spin (following Bardeen
1970;  where the counter-alignment or alignment conditions are taken
into consideration, i.e., BHs can be spun down or up), new $J_h$, new $R_w$, new $J_d$, new angle
between $J_h$ and $J_d$ (vectorial sum). In every timestep the disc
within $R_w$ is consumed.
\end{enumerate}

\begin{figure}
\includegraphics[width=0.9\columnwidth]{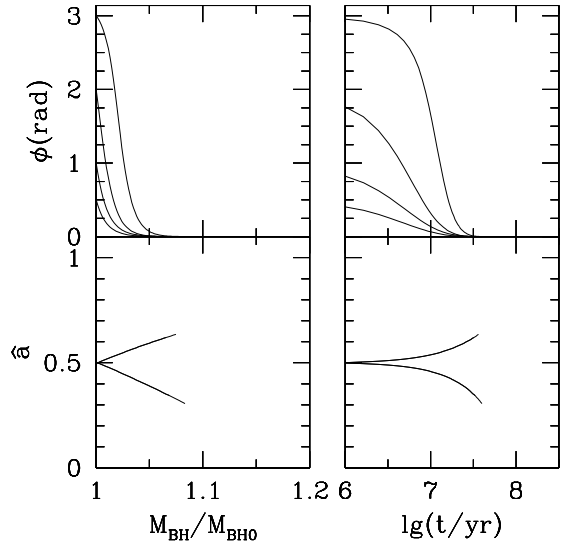}
\caption{Evolution of the misalignment angle between the angular momentum vector of the outer accretion disc and the BH spin (top panel), and of
the magnitude of the BH spin (bottom panel), due to accretion of
aligned material (spin-up) or counter-aligned material (spin-down).
The initial BH mass is $M_{\rm BH0}=5\times10^6 M_{\odot}$, the
initial spin $\hat a=0.5$, $\nu_2=\nu_1$, $\alpha=0.1$, and the
accretion rate is $f_{\rm Edd}=1$. The four curves show different
initial misalignment angles (top to bottom: $\phi=3, 2, 1, 0.5$
radians. ) }
\label{fig1}
\end{figure}

\begin{figure}[h]
\includegraphics[width=0.9\columnwidth]{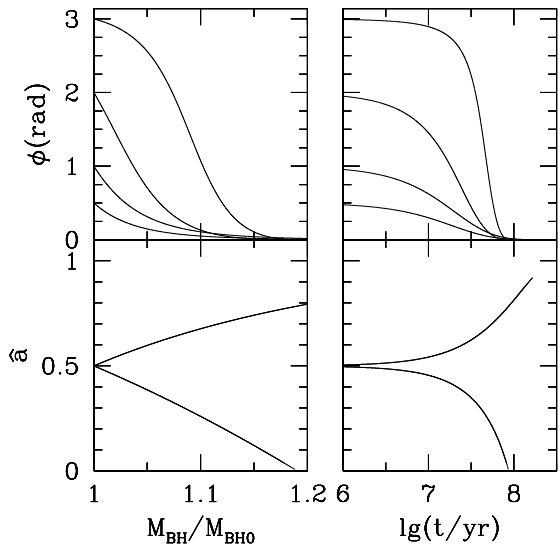}
\caption{As in Figure \ref{fig1}, but $\nu_2/\nu_1=1/\alpha^2$,
$\alpha=0.1$.}
\label{fig2}
\end{figure}

Figures \ref{fig1}  and \ref{fig2}  give examples of single accretion
episodes
for different initial angles between the angular momentum vector of the outer
(not warped) portions of the accretion disc and the black hole spin.
 They show the evolution of the spin magnitude and inclination as computed
for $\nu_2=\nu_1$ (Fig.1) and $\nu_2/\nu_1=1/\alpha^2$ (Fig.2).

%Here the
%initial BH mass is $M_{\rm BH}=5 \times 10^6 M_{\odot}$, and the initial spin
%$\hat a=0.5$.
% Black hole mass, initial spin and $\alpha$ are kept fixed
%for every curve, while the initial angle between outer disc and BH spin is
%varied - from initial counter-alignment to initial alignment. In
%Figure  \ref{fig1}: $\nu_2=\nu_1$, in Figure \ref{fig2}:
%$\nu_2/\nu_1=1/\alpha^2$. The calculations are stopped when $\phi=0$.

As we can see, the alignment timescale is basically independent of the
misalignment angle. To modify significantly the BH spin, one has to
bring to $R_w$ an amount of angular momentum comparable to $J_h$.
Therefore, if $J_h > J_d(R_w)$, then:
\begin{equation}
t_{\rm align}\simeq\frac{J_h}{J_d(R_w)}t_{\rm acc}(R_w)
\label{eq:talign1}
\end{equation}

Since $J_h\propto \hat a\,M_{\rm BH}\sqrt{R_S}$, $J_d(R_w)\propto
M_d(R_w)\sqrt{R_w}$, and
$M_d(R_w)=\dot M\,t_{\rm acc}(R_w)$, Equation \ref{eq:talign1} gives
(Rees 1978):
\begin{equation}
t_{\rm align}\simeq\hat a\frac{M_{\rm BH}}{\dot
M}\left(\frac{R_s}{R_w}\right)^{1/2}.
\label{eq:talign}
\end{equation}

Defining the mass accreted during $t_{\rm align}$ as $m_{\rm
align}=t_{\rm align}\dot M$, one gets:
\begin{equation}
m_{\rm align}\simeq M_{\rm BH}\,\hat a \left(\frac{R_s}{R_w}\right)^{1/2}.
\label{eq:malign}
\end{equation}
Therefore a series of many randomly oriented accretion events with
accreted mass $m \ll m_{\rm align}$ should result in black-hole's spin oscillating
around zero. For the opposite case of $m \gg m_{\rm align}$ the
black hole will be spun-up to large positive spins; for $m \sim
M_{\rm BH}$ the hole will be spun-up to $\hat a \sim 1$. Let us
notice finally that since it is reasonable to assume that $m_{\rm
align} \ll M_{\rm BH}$ the existence of AGN hosting black holes with
$\hat a \sim -1$ is rather unlikely.

Our calculations show that it is difficult to avoid high spin for the most massive black holes.
For large BH masses, the accretion timescale is very long, and
consequently the warp radius and the angular momentum within the
warp radius, $J_d$, are large. If $J_d>2\,J_h$, then the value
$|0.5\,J_d/J_h|>1$ and the counter-alignment condition cannot be
satisfied for any angle. This condition corresponds to:
\begin{eqnarray}
M_{\rm BH,max}>6.2\times 10^8 \msun\,
\alpha^{28/23}\times \nonumber \\ 
\left(\frac{\nu_2}{\nu_1}\right)^{19/23}
f_{\rm Edd}^{-2/23}\,\hat a^{-3/23}.
\label{eq:maxmass}
\end{eqnarray}
If $\nu_2=\nu_1$, $M_{\rm BH,max}$ is of order $10^7-10^8 \msun$ for most sensible choices of $\alpha$ and $f_{\rm Edd}$.  In this case the most massive black holes force accretion to occur from aligned discs, therefore causing a systematic spin-up. If the warp propagation is instead better described by $\nu_2=\nu_1/\alpha^2$, $M_{\rm BH,max}$  becomes exceedingly high and large accretion events can still act to spin down the black hole, provided $m<m_{\rm align}$.

The condition expressed in Equation \ref{eq:maxmass} is true only if there is enough mass to fill
the warp radius, that is if the total mass of the disc is larger than:
\begin{eqnarray}
M_{\rm d,min}>6.5\times 10^5 \msun\, \left(\frac{M_{\rm BH}}{10^8
\msun}\right)^{19/8}\alpha^{-3/2}\times \nonumber \\ 
f_{\rm Edd}^{1/4}\,\hat
a^{7/8}\left(\frac{\nu_2}{\nu_1}\right)^{-7/8}.
\label{eq:maxdisc}
\end{eqnarray}
If the mass to be accreted by the BH in an episode is smaller than $M_{\rm d,min}$,
then $J_d\ll J_h$, and both alignment or counter-alignment can happen.

\subsection{Multi-accretion events}
We then run a series of simulations in which we explore different
parameters. We start with a small BH, $M_{\rm BH0}=10^5 \msun$, and
have it grown by a series of accretion episodes. The accreted mass
$m$ is randomly extracted from only one of two different
distributions:
(1) a distribution flat in $m$, with $m<0.1M_{\rm BH}$, (2) a
distribution flat in $m$, with $m<0.01M_{\rm BH} $. The angle $\phi$ is
extracted from
a flat distribution $0<\phi<\pi$ at the beginning of every accretion episode.
Every simulation is composed by a large number of accretion episodes,
until one of the following conditions are met: $M_{\rm BH}>10^9\msun$ or $t_{\rm tot}>10^{10}$ years, that is the total simulation
time (total time a BH accretes to grow from the initial $M_{\rm
BH0}=10^5 \msun$ mass to its final mass) is shorter than the age of
the universe. During an episode where the BH accretes counter-aligned
material, the BH is spun down. If the black hole is spun-down until its spin
is zero,
any subsequently  accreted matter acts to spin the BH up again,
although the direction of the spin axis is now reversed
and aligned with the angular momentum of the disc.

\begin{figure}[h]
\includegraphics[width=0.9\columnwidth]{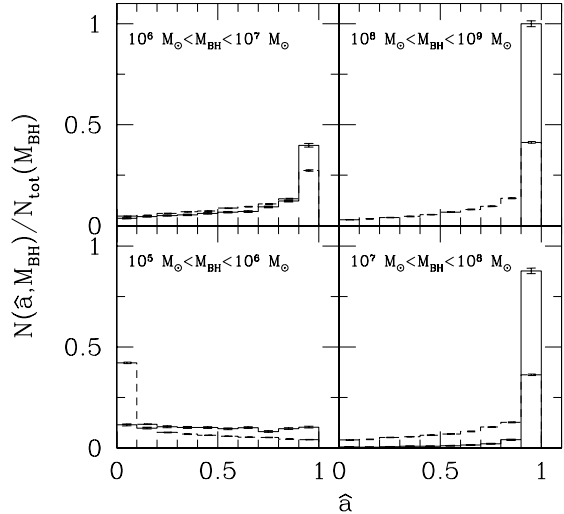}
\caption{Distribution of BH spins in different mass ranges. The accreted mass
$m$ is randomly extracted from a distribution flat in $m$, with
$m<0.1M_{\rm BH}$. Initial mass $M_{\rm BH0}=10^5 \msun$ , initial spin
$\hat a=10^{-3}$, $\alpha=0.03$, $f_{\rm Edd}=0.1$. Solid line: $\nu_2=\nu_1$,
dashed line: $\nu_2=\nu_1/\alpha^2$.}
\label{LM}
\end{figure}

We run 100 simulations for
every parameter sets choice, and we trace the
spins at the end of every accretion episode, for all the
accretion episodes in the simulations.

%In Figures \ref{fig5}, \ref{fig6}, \ref{fig7} we show the
%distribution of BH spins for a statistical ensemble.

We have explored a wide range of parameters, and we summarize here
our findings. We have varied the accretion rate, from $f_{\rm
Edd}=0.05$ to $f_{\rm Edd}=1$. If the accretion rate is low, the main caveat
is that black holes do not reach high masses within the Hubble time, however,
the efficiency of alignment is not strongly dependent on $f_{\rm Edd}$
(see Equation \ref{eq:malign}).

We have considered different  black hole spins at birth, from
$\hat a=10^{-3}$ to $\hat a=0.9$. After the BHs have changed their
initial mass by about one order of magnitude, the distributions are
indistinguishable from each other. During the first few e-foldings, however,
the spin distribution is  peaked around the black hole spin at birth.

We have also varied the viscosity parameter $\alpha$ (see section
2.1), and the relation between the viscosity characterizing the warp
propagation ($\nu_2$) with respect to the viscosity responsible for
the transfer of the component of the angular momentum parallel to
the spin of the disc ($\nu_1$). When $\alpha$  is varied, but
$\nu_2/\nu_1$ is kept fixed, the differences between the spin
distributions are not large. A smaller $\alpha$  skews the
distribution towards higher spins (cf Equations \ref{eq:rw},
\ref{eq:malign}).
\begin{figure}[h]
\includegraphics[width=0.9\columnwidth]{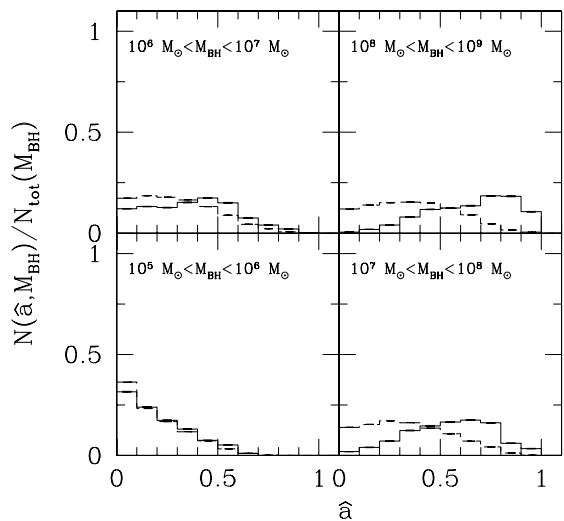}
\caption{ As in Figure \ref{LM}, but with
$m$ randomly extracted from a distribution flat in $m$, with
$m<0.01M_{\rm BH}$.}
\label{LLM}
\end{figure}

One of the main parameters influencing the spin distribution is the
relation between $\nu_2$ and $\nu_1$. If $\nu_2/\nu_1=1$, after the
BHs have reached $m_{\rm BH}\sim 10^6\msun$, the spin distribution
is dominated by rapidly spinning black holes. Equation
\ref{eq:maxmass} also shows that the most massive black holes force
accretion to occur from aligned discs, therefore causing a
systematic spin-up, unless very small parcels of material are
accreted at every single accretion episode. If the warp propagation
is instead better described by a high $\nu_2=\nu_1/\alpha^2$,
$M_{\rm BH,max}$ becomes exceedingly high, and all sorts of
accretion events can still act to spin down the black hole, provided
$m<m_{\rm align}$.

In fact, we confirm the results by MSL, that is that the main
parameter governing the distribution of BH spins is the amount of
material accreted in a single accretion episode.  This result is
clear from Figures \ref{LM}, \ref{LLM}  which refer
to different choices for the distributions of $m$. Only if
the mass accreted in one episode is smaller than $m_{\rm align}$,
the distribution of black hole spins can remain flat.

In the next section we discuss the likelihood of different $m$
distributions in the light of evolutionary  models for the BH
population in a hierarchical cosmology.

\subsection{Merger driven accretion}
\begin{figure}[h]
\includegraphics[width=0.9\columnwidth]{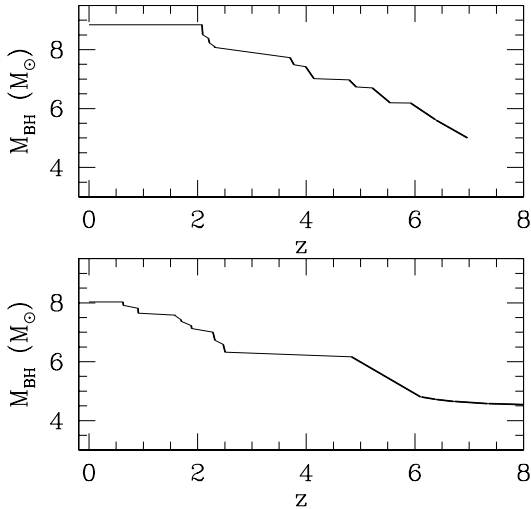}
\caption{Growth of BHs in  putative {\it "elliptical"} (upper panel)
or {\it "disc"} (lower panel) galaxy ( only merger driven accretion
events are considered).
The initial spin $\hat a=10^{-3}$, $\nu_2=\nu_1/\alpha^2$,
$\alpha=0.03$, and the accretion rate is $f_{\rm Edd}=0.1$.}
\label{evtrack}
\end{figure}

We first present an evolutionary track for BH spins, where a BH
grows by a sequence of randomly oriented accretion episodes in a
merger driven scenario. The BH mass evolutionary tracks are
extracted from semianalytical simulations of BH growth that have
been shown to reproduce the evolution of the BH population as traced
by the luminosity function of quasars (Marulli et al. 2006, 2007;
Volonteri, Salvaterra \& Haardt 2006). We focus here on two specific
tracks, for a putative BH in an {\it ``elliptical"}  (E) galaxy, and
one in a putative {\it ``disc"} (D) galaxy (Fig \ref{evtrack}).
Here the morphological classification is purely based on the
frequency of major mergers, i.e., mergers between comparable mass
galaxy systems which are believed to contribute mainly to the
spherical component of galaxies. A BH hosted in an {\it
``elliptical"} galaxy should have experienced a major accretion
event in connection with the last high-redshift major merger, which
formed the elliptical galaxy as we see it now. Afterwards, the galaxy (BH) has
not grown in mass due to merger driven star formation (accretion).
In the case of the BH hosted in a {\it ``disc"} galaxy, a small
number of minor mergers might have happened after the last major
mergers. These minor mergers are believed to be responsible for
re-building the galaxy disc. In conjunction  with these minor
mergers, a small infall of gas can produce a relatively minor
accretion episode onto the BH as well.

Along the evolutionary tracks for our BHs, we trace the joint evolution of accretion onto the BH, the
dynamics of the accretion disc, and the consequences on the spin.
The scheme we adopt is similar to the one described in Section 4.1.
During an episode where the BH accretes counter-aligned material, the
BH is spun down, until the spin is zero, and subsequently any
accreted matter acts to spin the BH up again, although the direction
of the spin axis is now reversed and aligned with the angular
momentum of the disc.

In case ``E",  the last accretion episode  caused a large
increase in the BH mass, following the major merger which created the
elliptical itself (Hopkins \& Hernquist 2006).  During this episode
the spin increased significantly as well, up to very high values.
Let us remind here that in the extreme event of a maximally-rotating
hole spun down by retrograde accretion, the BH is braked after
its mass has increased by the factor $\sqrt{3/2}$. Any mass
accreted afterwards  spins up the black hole,  and if the final
mass increase is by a factor 3, the BH will end up
maximally-rotating again.

Let us now consider a ``D"-type evolution. A BH would experience a
series of small accretion episodes (triggered possibly by minor mergers),
extending for a longer period of times.  If these episodes are
uncorrelated, that is if the inflow during a given episode is not
aligned with the orientation of the spin of the BH, the
randomization of the angle $\phi$ over the (few) accretion episodes
tends to spin down the BH.

\begin{figure}
\includegraphics[width=0.9\columnwidth]{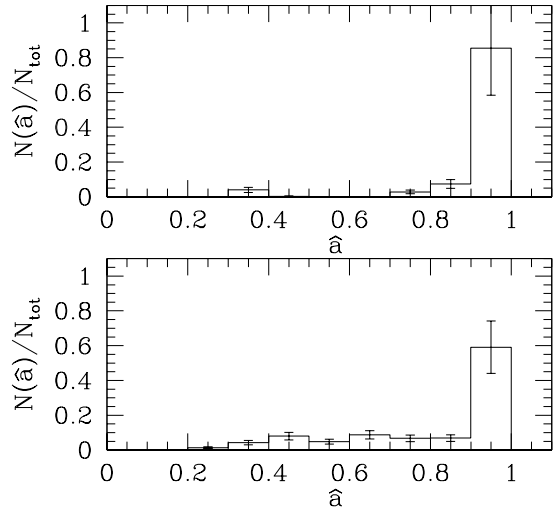}
\caption{Spin of BHs at $z=0$ in {\it "elliptical"} (upper panel)
or {\it "disc"} (lower panel) galaxies.}
%The initial spin $\hat a=10^{-3}$, $\nu_2=\nu_1/\alpha^2$, $\alpha=0.03$, and
%the accretion rate is $f_{\rm Edd}=0.1$.
\label{evtrack1}
\end{figure}

We run a statistical sample of ``E" and ``D" track, for  BHs hosted
in large (i.e., Andromeda-size systems) galaxies. We find that, if
only the merger driven evolution is taken into account,  BHs in {\it
``elliptical"} galaxies are left with large spins. BHs in {\it
``disc"} galaxies have, on average, slightly lower spins, however
the distribution is still peaked at large values (Figure \ref{evtrack1}).

\subsection{Short-lived accretion events}
Even minor mergers tend to trigger inflows of matter which are too
large to  lead to the series of short lived accretion events
necessary to leave BHs with small spins (cfr. the discussion in
MSL). Moreover, several observations suggest that single accretion
events last  $\simeq 10^5$ years in Seyfert galaxies, while the total activity lifetime (based on the fraction of disc galaxies that are Seyfert) is
$10^8-10^9$ years (e.g., Kharb et al. 2006; Ho et al. 1997). This
suggests that accretion events are very small and very
{`}compact'.

A type of random event which leads to short-lived accretion episodes
is the tidal disruption of stars. One expects discs formed by
stellar debris to form with a random orientation. Stellar
disruptions would therefore contribute to the spin-down of BHs. Let
us consider the maximal influence that feeding via tidal disruption
of stars can have on spinning down a BH. The number of tidal
disruptions of solar type stars in an isothermal cusp per billion
years can be written as:

\be N_*=4\times 10^5
\left(\frac{\sigma}{60\, {\rm km/s}}\right)\left(\frac{M_{\rm
BH}}{10^6 \msun}\right)^{-1}
\ee.

Assuming that BH masses scale with
the velocity dispersion, $\sigma$, of the galaxy (we adopt here the
Tremaine et al. 2002 scaling), we can derive the relative mass
increase for a BH in 1 billion years:
\be \frac{M_*}{M_{\rm
BH}}=0.37\left(\frac{M_{\rm BH}}{10^6\msun}\right)^{-9/8}.
\label{eqTD}
\ee
The maximal level of spin down would occur assuming
that all the tidal disruption events form counterrotating discs,
leading to retrograde accretion (note that the mass of the debris disc is much smaller than $M_{\rm d,min}$, cfr. Eq. \ref{eq:maxdisc}, so that counter-alignment is allowed for any BH mass). Eq. \ref{eqTD} shows that a small (say $10^6 \msun$) BH starting at $\hat a=0.998$ would be spun down completely, on the other hand the spin of a larger (say $10^7 \msun$) BH would not be changed
drastically.

Early type discs typically host faint bulges characterized by steep
density cusps, both  inside (Bahcall \& Wolf 1976; Merritt \& Szell
2006) and outside (Faber et al. 1997) the sphere of influence of the
BH.  In this environment, the rate of stars which are tidally
disrupted by BHs  (Hills 1975; Rees 1988)  less massive than
$10^8\msun$\footnote{For black hole masses $\geqslant 2\times
10^8\msun$ the Schwarzschild radius exceeds the tidal disruption
radius for main-sequence stars.} is non negligible (Milosavljevic et
al. 2006). Since in elliptical galaxies the
central relaxation timescale is typically longer than the Hubble
time, and the central density profile often displays a shallow core,
tidal disruption of stars is unlikely to play a dominant role.

An additional feeding mechanism might be at work in gas-rich
galaxies with active star  formation. Compact self-gravitating cores
of molecular clouds (MC) can occasionally reach subparsec regions,
and may do it with random directions provided that the galactic disc
is much thicker than the spatial scale of the BH gravity domination
region (Shlosman, private communication).
Although the rate of such events is uncertain, we can adopt
the estimates of Kharb et al. (2006), and assume that about $10^4$ of
such events happen. We can further assume a lognormal distribution
for the mass function of MC close to galaxy centers (based on the
Milky Way case, e.g., Perets, Hopman \& Alexander 2006). We do not
distinguish here giant MC and clumps, and, for illustrative purpose
we assume a single lognormal distribution peaked at
$\log(M_{\rm MC}/\msun)=4$, with a dispersion of 0.75.

\begin{figure}
\includegraphics[width=0.9\columnwidth]{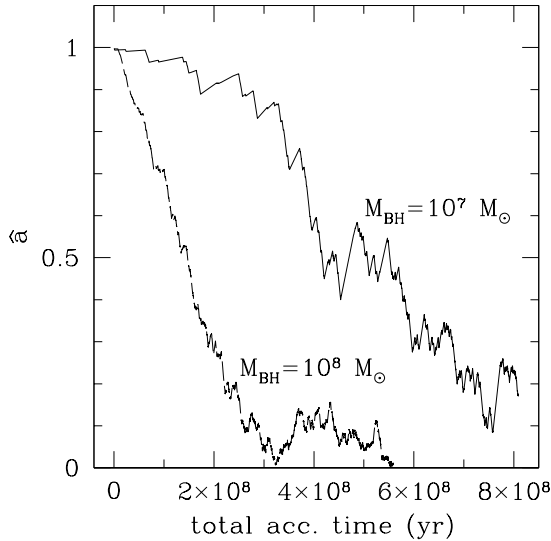}
\caption{Evolution of BH spins due to accretion of molecular clouds
cores. We assume a lognormal distribution for the mass function of
molecular clouds (peaked at $\log(M_{MC}/\msun)=4$, with a dispersion
of 0.75. The initial spin of the BHs is 0.998. Upper curve: the
initial BH mass is $10^7\msun$, lower curve: the initial BH mass is
$10^8\msun$}
\label{fig6}
\end{figure}

Fig \ref{fig6} shows the possible effect that accretion of molecular
clouds  can have on spinning BHs. The result is, on the whole,
similar to that produced by minor mergers of black holes (Hughes \& Blandford
2003), that is a spin down in a random walk fashion.  The larger
the BH mass, the more effective the spin down.

In a gas-poor elliptical galaxy, however,
substantial populations of molecular clouds are lacking (e.g. Sage et al. 2007),
thus hampering the latter mechanism for short lived accretion events
proposed.

\section{Discussion and conclusions}

We have investigated the evolution of BH spins driven by accretion
from discs with angular momentum vectors that can be misaligned with respect to the
spin axis.
We have assumed that
accretion discs can be described by Shakura--Sunyaev $\alpha$-discs,
and that when the angular momentum of the accretion disc is not
aligned with the spin of the BHs, the disc itself is warped. The
inner portions of the discs experience Lense-Thirring torque,
which tends to align the inner parts of the disc. The timescale of the
Lense-Thirring precession increases faster with distance from a BH than the timescale of
warp propagation, and they equate  at the so-called ``warp radius",
where a transition occurs from alignment to misalignment.  King et
al. (2005)
pointed out that for highly misaligned discs, counter-alignment,
rather than alignment
can occur. The co- or counter-alignment of the accretion discs has
important consequences on the spin of BHs. A black hole accreting
from prograde orbits (i.e., alignment case) is spun up by the
coupling between the angular momentum of the infalling material and
its spin (Bardeen 1970). If, instead, an initially spinning hole
accretes from retrograde orbits (i.e., counter-alignment case), it
is spun down. An initially non-rotating BH gets spun up to a
maximally-rotating  state ($\hat a=1$) after  reaching the mass
$M_{\rm BH}=\sqrt{6}M_{\rm BH0}$. A
maximally-rotating hole ($\hat a=1$) gets spun down by retrograde
accretion  to $\hat a=0$ after reaching the mass
$M_{\rm BH}=\sqrt{3/2}M_{\rm BH0}$.
A $180^\circ$ flip of the spin of an extreme-Kerr  hole
will occur after $M_{\rm BH}=3M_{\rm BH0}$.

It is therefore necessary that accretion episodes increase the mass
of a BH by less than $M_{\rm BH}=\sqrt{6}M_{\rm BH0}$, in
order to keep the spin at low values, if accretion preferentially
occurs from prograde orbits. Natarajan \& Pringle (1998) suggested
that accretion indeed occurs from aligned discs (i.e. prograde orbits),
as the timescale for disc alignment is
much shorter than the timescale  of the BH mass growth by
$M_{\rm BH}=\sqrt{6}M_{\rm BH0}$.
King et al. (2005) suggested, however,
that when the initial misalignment angle is large and $m$ is
sufficiently small, counter-alignment, rather than alignment, occurs
and BHs can be spun down in a large fraction of the accretion
episodes.

We have quantified here the likelihood of counter-alignment and spin
down as claimed by King et al. (2005).  We identify two  main
parameters influencing the distribution of BH spins: the
distribution of the accreted mass,  $m$, with respect to the mass of
the BH, and the relation between $\nu_2$ and $\nu_1$, where $\nu_2$
is the viscosity characterizing the warp propagation, and $\nu_1$,
which is responsible for the transfer of the component of the
angular momentum parallel to the spin of the disc. $\nu_2$ can in
principle differ from $\nu_1$.

If the accreted mass, $m$, is much smaller than the mass of the BH
(e.g., $m<0.01\, M_{\rm BH}$), the distribution of black hole spins
is flat, as the timescale for spin overflipping due to the
Bardeen--Petterson effect is longer than the timescale to accrete
the whole $m$. If instead $m\simeq M_{\rm BH}$, BHs can align with
the angular momentum of the accretion disc, and accrete enough mass
to be spun up. In this case the distribution of BH spins is
dominated by rapidly rotating systems.

Understanding if  the description of the warp propagation is
correctly described by a different viscosity with respect to the one
responsible for the radial propagation of the angular momentum is
beyond the scope of this work. We have therefore explored a wide
range of possible viscosities, and we simply report here our
results.  If $\nu_2/\nu_1=1$ the timescale for alignment is short,
and the spin of a BH increases rapidly. If the warp propagation is
instead better described by a high $\nu_2=\nu_1/\alpha^2$, a
substantial fraction of black holes of all masses can have small
spins, provided $m\ll M_{\rm BH0}$.

However, both semi-analytical models of the cosmic BH evolution
(Volonteri et al. 2005) and simulations of merger driven accretion
(di Matteo et al. 2005) show that most BHs increase their mass by
an amount $\gg m_{\rm align}$, if the evolution of the LF of quasars
is kept as a constraint. These high $m$ values are likely
characteristic of the most luminous quasars and most massive black
holes -- especially at high redshift. We expect therefore that
bright quasars at $z>3$ have large spins, in contrast with the
suggestion of King \& Pringle (2006).  High spins in bright quasars
are also indicated by the high radiative efficiency of quasars, as
deduced from observations applying the So{\l}tan argument (So{\l}tan
1982; Wang et al. 2006 and references therein).

If  the mass of a BH need to reach $10^9\msun$ by $z=3$, or
even more strikingly, by $z=6$, so that they can represent the
engines of quasars with luminosity $L>10^{46}\rm{erg \, s^{-1}}$, BHs
need to grow from typical seed masses (e.g. Madau \& Rees 2001,
Koushiappas et al. 2003, Begelman, Volonteri \& Rees 2006, Lodato \&
Natarajan 2006) by at least 3-4 orders of magnitude in $10^8-10^9$
years. The necessity of  long and continuous accretion
episodes implies therefore that, for these BHs,
$m\gg m_{\rm align}$.

Smaller BHs, powering low luminosity Active Galactic Nuclei, can
instead grow by accreting smaller packets of material, such as
tidally disrupted stars (for BHs with mass $<2\times 10^6 \msun$,
Milosavljevic et al. 2006), or possibly molecular clouds (Hopkins \&
Hernquist 2006).  For these black holes the spin distribution is
more probably flat, or skewed towards low values. This latter result
is in agreement with Sikora, Stawarz \& Lasota (2006), who find that
disc galaxies tend to be weaker radio sources with respect to
elliptical hosts. In the hierarchical framework we might expect that
the BH hosted by an elliptical galaxy had, as last major accretion
episode, a large increase in its mass following the major merger
which created the elliptical itself (Hopkins \& Hernquist 2006).
During this episode the spin increased significantly as well,
possibly up to very high values. Subsequently the black hole might
have grown by swallowing the occasional molecular cloud, or by
tidally disrupting stars. If the total contribution of these random
episodes represents a small fraction of the BH mass, the spin is,
however, kept at high values.

Black holes in spiral galaxies, on the other hand, probably had
their last major merger (i.e., last major accretion episode), if
any, at high redshift, so that enough time elapsed for the galaxy
disc to reform. Most of the latest growth of the BH should have
happened through minor events, which have likely contributed to the
BH spin down.

Our results are supported also by the recent finding by Capetti
\& Balmaverde (2006; 2007)  that radio bimodality correlates with
bimodality of stellar brightness profiles in galactic nuclei. The
inner regions of radio loud galaxies display shallow cores (star
deficient). Cores, in turn,  are preferentially  reside in giant
ellipticals (see Lauer et al. 2007 and references therein). Radio
quiet galaxies, including nearby low luminosity Seyferts, have
instead power-law (cuspy) brightness profiles and preferentially
reside in SO and spiral galaxies.

Hence, noting that core nuclei result from  merging BHs  following
galaxy mergers (Ebisuzaki et al. 1991, Milosavljevic \&
Merritt 2001, Milosavljevic et al. 2002, Ravindranath et al. 2002,
Volonteri et al. 2003), Balmaverde \& Capetti's discovery is
consistent with our conjecture that spin bimodality is  determined
by diverse evolutionary tracks of BH spins in disc galaxies (random
small mass accretion events) and giant elliptical galaxies (massive
accretion events which follow galaxy mergers). Both tidal disruption
of stars, and accretion of gaseous clouds is unlikely in shallow,
stellar dominated galaxy cores. Therefore it is conceivable that the
observed morphology-related bimodality of AGN radio-loudness results
from bimodality of central black-holes spin distribution.

\acknowledgements M.S. thanks the Dept. of Physics and Astronomy, University of Kentucky at Lexington, for hospitality during his stay. This research was
supported in part by the National Science Foundation under Grant No.
PHY99-07949.

%\bibliographystyle{mn2e}
%\bibliography{xrc.bib}

\end{document}